\documentclass[3p]{elsarticle}
\usepackage{verbatim}
\usepackage{geometry}
\usepackage[T1]{fontenc}
\usepackage[utf8]{inputenc}
\usepackage[american]{babel}
\usepackage{epsfig}
\usepackage{graphicx,subcaption,caption}
\usepackage{booktabs}
\usepackage{multirow}
\usepackage{dcolumn}
\usepackage{amsmath}
\usepackage{mathtools}
\usepackage{amsfonts}
\usepackage{amssymb}
\usepackage{epstopdf}
\usepackage{bm}
\usepackage{siunitx}
\usepackage{braket}
\usepackage{enumitem}
\usepackage{soul}
\usepackage[table]{xcolor}
\usepackage{color}
\usepackage{transparent}
\usepackage{pifont}
\biboptions{sort&compress}

\definecolor{navyblue}{rgb}{0.0, 0.0, 0.5}
\definecolor{royalblue}{rgb}{0.25, 0.41, 0.88}
\definecolor{cadmiumgreen}{rgb}{0.0, 0.42, 0.24}
\definecolor{blue-violet}{rgb}{0.54, 0.17, 0.89}
\definecolor{darkviolet}{rgb}{0.58, 0.0, 0.83}
\definecolor{orange(colorwheel)}{rgb}{1.0, 0.5, 0.0}

\usepackage{hyperref}
\hypersetup{
    colorlinks=true, 
    linkcolor=royalblue, 
    citecolor=magenta}

\usepackage{booktabs}
\usepackage{multirow}
\usepackage{dcolumn}
\usepackage{colortbl}

\begin{document}

\begin{frontmatter}

\title{Probing the inflationary background of gravitational waves from large to small scales}

\author[1]{William Giarè \corref{CorrAuthor}}
\ead{william.giare@uniroma1.it}

\author[1]{Alessandro Melchiorri}
\ead{alessandro.melchiorri@roma1.infn.it}

\address[1]{Physics Department and INFN, Universit\`a di Roma ``La Sapienza'', Ple Aldo Moro 2, 00185, Rome, Italy}

\cortext[CorrAuthor]{Corresponding author}

\date{\today}
\begin{abstract}
The detection of Primordial Gravitational Waves (PGWs) is one of the most important goals of modern cosmology since PGWs can both provide substantial evidence for primordial inflation and shed light on its physical nature. 
Small scale experiments on gravitational waves such as LIGO/VIRGO and, in future, LISA and Einstein Telescope (ET), being sensitive to the stochastic background of gravitational waves, can be used together with the CMB data to constrain the inflationary parameters. In performing these analyses the primordial tensor spectrum is usually parametrized with a power law that includes only the amplitude and a scale independent tilt.
In this paper, we investigate the robustness of assuming the tensor tilt as scale independent. We show that due to the huge difference in the scales probed by CMB and GWs data, even a small scale dependence can remarkably affect the shape of the primordial spectrum possibly breaking the power-law assumption. When the non-linear corrections are considered the final constraints can be significantly changed. We also study the scale dependence in two different physical models of inflation providing an example of negligible scale dependence and an example of non-negligible scale dependence.
\end{abstract}

	
\end{frontmatter}


\section{Introduction} \label{sec.Intro}
The search for primordial gravitational waves (PGWs) is one of the main goals of modern cosmology as they can both provide a substantial evidence for primordial inflation and shed light on its physical nature \cite{Kamionkowski:2015yta,Baumann:2014cja,Caldwell:2018giq,Franciolini:2018ebs}. In fact long-wavelength gravitational waves are predicted by primordial cosmic inflation \cite{Guth:1980zm,Starobinsky:1980te,Linde:1981mu,Vilenkin:1983xq} and, at least in the simplest models, the scale at which inflation occurs is itself related to the amount of PGWs \cite{Sahni:1990tx, Lyth:2009zz,Mukhanov:2005sc,Dodelson:2003ft,Weinberg:2008zzc,Martin:2013tda,Kamionkowski:2015yta,Mirbabayi:2014jqa,Ozsoy:2014sba}. The missing evidence of the B-modes in the Cosmic Microwave Background (CMB) polarization originated from the inflationary tensor modes and, in general, a combined analysis of the Planck and BICEP2/Keck array (BK15) data \cite{Ade:2018gkx}, allow us to set only an upper bound on the amplitude of PGWs on the CMB scales $r<0.07$ at 95\% C.L. at the pivot scale $k_* = 0.05\,\rm{Mpc^{-1}}$ \cite{Akrami:2018odb}. Nevertheless, in the upcoming decade, a new generation of CMB experiments such as BICEP3 \cite{BICEP3}, CLASS \cite{CLASS} , SPT-3G \cite{SPT-3G}, Advanced ACTPol \cite{ACTPol}, LBIRD \cite{LBIRD} and CMB-S4 \cite{CMB-S4}  are expected to bring the sensitivity to the tensor amplitude down to $r \sim 0.01 - 0.001$ possibly leading to its first detection.

In deriving such bounds the slow roll consistency relation $n_{\rm t}=-r/8$ is usually assumed, basically leading to an almost scale independent slightly red tilted primordial tensor spectrum $\mathcal P_{\rm t} (k)$. However this relation can be violated in many non standard models of inflation and when it is relaxed the Planck data only weakly constrain the tensor tilt to $-0.55<n_{\rm T}<2.54$ at 95\% C.L. \cite{Akrami:2018odb}. Nevertheless, always in \cite{Akrami:2018odb}, it was shown that a significant improvement in the upper bound on the tensor tilt can be obtained considering the LIGO/VIRGO data on the stochastic background of gravitational waves $\Omega_{\rm GW}$, \textit{i.e.} the analogous of CMB for gravitational waves \cite{Caprini_2018}. Indeed, while a direct detection of the stochastic background has not been provided yet, in the frequency range $f\in\left(20\,\rm{-}\,85.8\right)$ Hz, which corresponds to the wavenumber range  $k \in \left(1.3\,\rm{-}\,5.5\right) \times 10^{16} \,\rm{Mpc}^{-1}$, the first and second observing runs of the LIGO/VIRGO collaboration set an upper bound on the stochastic background
 \begin{equation}
 \Omega_{\rm{GW}} (k_{\rm LV}) \leq 1.7 \times 10^{-7} 
 \label{LVlimit}
 \end{equation}
at 95 \% C.L.  \cite{Akrami:2018odb,TheLIGOScientific:2016dpb}.
The Planck Collaboration, including the LIGO/VIRGO limit \eqref{LVlimit} as a half-gaussian prior on the tensor tilt, under the assumption of scale independence, derived the improved upper bound  $n_T < 0.53$ at 95\% C.L. \cite{Akrami:2018odb}. Notice that these constraints are obtained marginalizing over the probability distribution of $r$, that is typically sampled assuming a flat prior $r\in[r_{\min}\,,\,r_{\rm max}]$ with $r_{\rm min}\lesssim 10^{-3}\simeq 0$. This makes the constraint on the tensor tilt subject to misunderstanding as if there is no detection of $r$ no reliable constraint can be derived. Moreover, another important assumption beyond this analysis, is to consider the tensor tilt as scale independent extending the well known power law relation
\begin{equation}
\mathcal P_{\rm t}(k) = r\, A_{\rm s}\, \left( \frac{k}{k_{*}}\right)^{n_{\rm t}}
\label{no_runnings}
\end{equation}
(with $A_S = 2.1\times 10^{-9}$ the scalar amplitude at the pivot scale $k_*=0.05\rm{Mpc}^{-1}$) from the CMB scales ($k\sim 0.05 \, \rm{Mpc}^{-1})$ all the way up to the small scales probed by the gravitational interferometers ($k\sim 10^{16} \, \rm{Mpc}^{-1}$).  This is clearly an approximation as Eq. \eqref{no_runnings} is just a first order expansion: depending on the model of inflation, $n_{\rm t}$ can also acquire a slight scale dependence and non-linearities may break the power-law relation. It is therefore timely to investigate the impact of higher-order corrections on the constraints one can derive exploiting small-scale data on Gravitational Waves. We show that due to the huge difference in the scales probed by CMB and GW data, the higher-order terms in the spectrum, the so-called tensor runnings  \cite{Giare:2019snj,Kuroyanagi:2011iw}, albeit small on the CMB scales, may give non-negligible contributions on the ultrahigh $k$ probed by gravitational detectors, drastically changing the final predictions.

The paper is organized as follows: in Sec. \ref{sec.ScaleDependence} we discuss the constraints on the inflationary parameters from gravitational wave experiments allowing the possibility to have a scale dependent tilt. We show that the results strongly depend on the assumption of scale-independence. 
In Sec. \ref{sec.Examples} we study the scale dependence in two different physical models of inflation. In particular in Sec. \ref{sec.Starobinsky} we consider the Starobinsky model showing that it predicts a negligible scale dependence as expected in the standard slow roll paradigm. In Sec. \ref{sec.ParticleProduction} instead we study a more elaborated scenario of inflation that employs a pseudo scalar axion naturally coupled to gauge fields. We show that in this model the tensor tilt can acquire a non negligible scale dependence leading to appreciable corrections on small scales. 
In Sec. \ref{sec.Conclusion} we present our conclusion.

\section{Scale dependence} \label{sec.ScaleDependence}
\begin{figure}[h]
 	\centering
 	\includegraphics[width=0.75\linewidth]{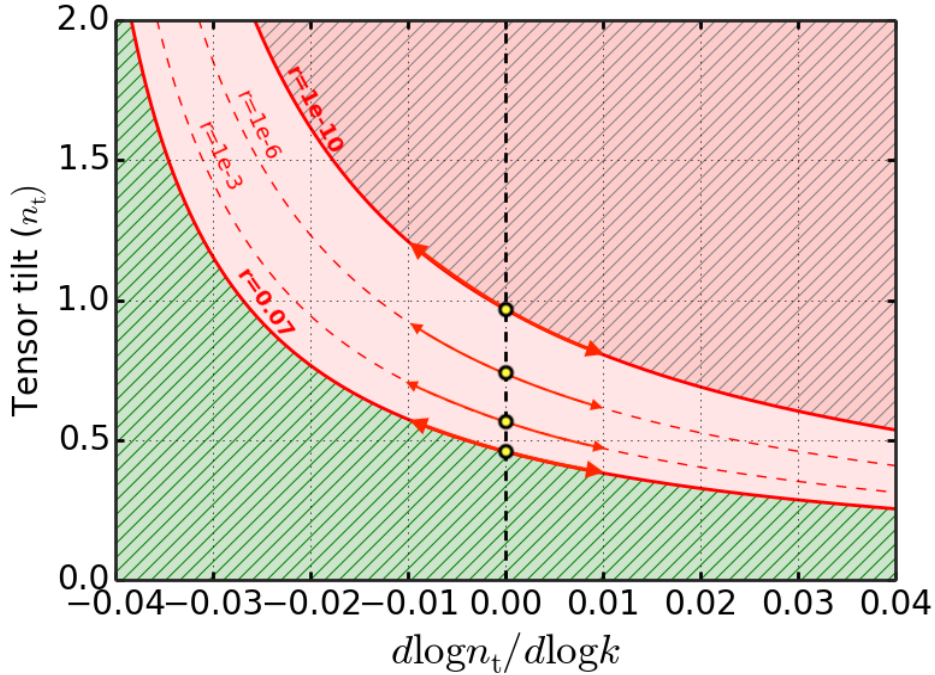}
 	\caption{Constraints on the tensor tilt from the LIGO/VIRGO limit on the stochastic background \eqref{LVlimit}. The yellow dots represent the upper bounds on $n_{\rm t}$ for different values of $r$ when scale-dependence is ignored. When a scale dependence $d\log n_{\rm t} / d\log k\ne0$ is considered the yellow dots move on the red lines at constant $r$.}
 	\label{fig:figure1}
 \end{figure}

Along with B-modes polarization, primordial tensor fluctuations may have imprinted also the stochastic background of gravitational waves, the analogous of CMB for gravitational waves \cite{Caprini_2018}. The energy density of the universe due to PGWs at the present time and at a given scale $k=2\pi\,f$ is given by \cite{Akrami:2018odb,Bartolo:2016ami,Cabass:2015jwe,Stewart:2007fu,Graef:2018fzu}
 \begin{equation}
 \Omega_{\mathrm{GW}}(k) \doteq \frac{1}{\rho_{c}} \frac{\mathrm{d} \rho_{\mathrm{GW}}}{\mathrm{d} \log k}\simeq \frac{\mathcal P_{\mathrm{t}} (k)} {24 z_{\mathrm{eq}}}
 \label{Omega_GW}
 \end{equation}
where $z_{\rm{eq}}\simeq 3400$ is the redshift at the matter-radiation equivalence \cite{Akrami:2018odb}. Using Eq. \eqref{no_runnings}, it is easy to see that, under the assumption of scale-independent tilt, a constraint on the amplitude of the stochastic background $\Omega_{\rm GW}(k)$ can be translated into an upper bound on the tensor tilt  
\begin{equation}
 n_{\rm t} <\frac{\ln\left(\frac{24\,z_{\rm eq}\,\Omega_{\rm GW}(k)}{r\,A_{\rm S}}\right)}{\ln\left(\frac{k}{k_*}\right)} \lesssim 0.39 + 0.025\times \log(1/r),
 \label{blue_tilt}
 \end{equation}
where in the last inequality we considered the LIGO/VIRGO limit \eqref{LVlimit}.
Note that the constraint \eqref{blue_tilt} is derived without any assumption on the tensor amplitude\footnote{This is a different approach with respect to those performed in \cite{Akrami:2018odb} where the upper bound $n_{\rm t}<0.53$ at 95\% C.L. was derived marginalizing over the distribution of $r$. Anyway we see that for $r\sim 10^{-2} - 10^{-3}$ we basically recover the same result, see also Fig. \ref{fig:figure1}.}: if future measurements will reveal evidence for $r \ne 0$, its detection will immediately place a well defined upper bound on the tensor tilt that however, because of its logarithmic dependence, is not drastically sensitive to the precise value of the tensor amplitude, as one can also appreciate in Fig. \ref{fig:figure1}. The physical reason beyond this weak dependence on the tensor amplitude is actually that a large positive tilt will strongly amplify the GWs production on the ultrahigh $k$ probed by gravitational detectors, easily compensating a small (but of course not vanishing) $r$ on the CMB scales. On the other hand, it is also true that constraints on $n_{\rm t}$ cannot be derived for a vanishing tensor amplitude, and in fact we may see that taking the limit $r\to 0$, the right side of Eq. \eqref{blue_tilt} logarithmically diverges, as expected.

As we said, the constraint \eqref{blue_tilt} as well as the upper bound  $n_{\rm t}<0.53$ at 95\% C.L. \cite{Akrami:2018odb} is derived assuming a constant tilt over a range of about eighteen order of magnitude, namely $k\in[0.05\,,\,1.3\times 10^{16}]\,\rm{Mpc^{-1}}$. In order to include a possible scale dependence, we generalize the power law parametrization to the following expansions:
 \begin{equation}
 \mathcal P_{\rm t}(k) =  r\, A_{\rm s}\,\left(\frac{k}{k_*}\right)^{n_{\rm t}(k_*) + \sum_{n=1}^{\infty} \frac{\alpha_n^{\rm t}(k_*)} {(n+1)!} \left[\log\left(\frac{k}{k_*}\right)\right]^n}
 \label{Tensor}
 \end{equation} 
 where 
 \begin{equation}
 \alpha_n^{\rm t}(k_*)\doteq \left( \frac{d}{d\log k}\right)^n\,n_{\rm t}(k)\bigg \rvert _{k=k_*}
 \end{equation}
 is the $n$-order tensor running of the tensor tilt\footnote{In what follows we will usually avoid to specify that the spectral tilt and the runnings are computed on the pivot scale $k_*$ and, to simplify the notation, we will only write $n_{\rm t}$ and $\alpha_n^{\rm t}$.}. When the runnings are included, the upper bound \eqref{blue_tilt} is modified to
 
 \begin{equation}
 n_{\rm t}\lesssim \frac{\ln\left(\frac{24\,z_{\rm eq}\,\Omega_{\rm GW}(k)}{r\,A_{\rm S}}\right)}{\log\left(\frac{k}{k_*}\right)}  -  \sum_{n=1}^{\infty} \frac{\alpha_n^{\rm t}(k_*)} {(n+1)!} \left[\log\left(\frac{k}{k_*}\right)\right]^n.
 \label{runnings}
 \end{equation}
 Clearly, in order to exactly compute the sum expansion and to check its convergence we need to estimate all the runnings $\{\alpha_{n}^{\rm{t}}\}$ and this is possible only fixing a specific model of inflation.  Nevertheless we can appreciate how the generic $n$-order running must at least satisfy the condition $|\alpha_n^{\rm t}/n_{\rm t}|\ll (n+1)! / \log^n(k/k_*)$
to give a negligible contribution at the generic scale $k$. This should become a non trivial requirement, above all on the ultra-high $k$ as those probed by gravitational interferometers.

To study how constraints on $n_{\rm t}$ derived under the assumption of scale independence are modified in presence of a slight scale dependence, we derive the upper bound on the tensor tilt by the LIGO/VIRGO limit \eqref{LVlimit} for different value of the tensor to scalar ratio, varying $d\log n_{\rm t} / d\log k$ - that represents the rate of change of the tensor tilt with respect to the scale - in a range $d\log n_{\rm t} / d\log k\in \left[-0.04\,,\,0.04\right]$. We find that, because of the huge difference between the scales probed by CMB and GW data, also a small departure from scale independence ($\lesssim 4\%$) can significantly change the final results, see also Fig. \ref{fig:figure1}.  In particular, a small negative (positive) running\footnote{We recall that $d\log n_{\rm t}/d\log k\doteq \alpha_1^{\rm t} / n_{\rm t}$.}, suppressing (amplifying) the amplitude of PGWs on small scales, can remarkably worsen (improve) the upper bound derived under the assumption $d\log n_{\rm t} / d\log k =0$ (yellow dots in Fig. \ref{fig:figure1}). This is clearly translated into a strong degeneracy between scale-dependence and the tensor amplitude (see Fig. \ref{fig:figure1}) that can be broken only by an independent measurement of $r$ from future CMB experiments.
We conclude that the small-scale constraints on $n_{\rm t} $ may be very sensitive to the assumption of scale independence: non-negligible contributions can arise from non-linear corrections and they cannot be always ignored when constraints on the inflationary parameters are derived exploiting GW data. 
\section{Examples} \label{sec.Examples}
To further validate our discussion, in this section we study two physical models of inflation. We first analyze the Starobinsky model that, being pure slow roll, by definition predicts an almost scale independent slightly red tilted spectrum. It represents an example of models where higher order corrections are typically negligible also on small scales. However this cannot be true in more elaborated scenarios: as a counterexample we study a model of particle production where non linear corrections lead to a non negligible scale dependence.
\subsection{Starobinsky Inflation}\label{sec.Starobinsky}
In the simplest framework of single field slow roll inflation one can compute the primordial spectra both for scalar and tensor perturbations to obtain \cite{Riotto:2018pcx,Martin:2013tda,Guth:1985ya,Mukhanov:1990me,Starobinsky:1979ty,Mukhanov:2013tua,Bartolo:2001rt}:
\begin{equation}
\mathcal{P} _ {\rm s } = \left(\frac{1}{8\pi^2 M_{p}^2}\right)\left(\frac{H^2}{\epsilon_1}\right)
\label{scalar_spectrum}
\end{equation}
\begin{equation}
\mathcal{P} _ {\rm t } = \left(\frac{2}{\pi^2 M_{p}^2}\right)H^2
\label{tensor_spectrum}
\end{equation}
where $M_{\rm p}$ is the reduced Planck mass ($=2.435\times10^{18}$ GeV), $H$ is the Hubble parameter, $\epsilon_1\doteq-\dot H /H^2$ is the first of the slow roll parameters $\{\epsilon_1\,...,\epsilon_n\}$ defined as $\epsilon_{n>1}\doteq d\log\epsilon_{n-1} /d\log k$. All these relations are to be considered calculated at the horizon crossing.
We also introduce the usual slow roll relations for the tensor to scalar ratio $r$, the tensor tilt $n_{\rm t}$ and the scalar tilt $n_{\rm s}$ \cite{Martin:2013tda,Giare:2019snj}:
\begin{equation}
n_{\rm t}\doteq \frac{d\log \mathcal P_{\rm t}}{d\log k}=-2\epsilon_1 =- \frac{r}{8}
\label{nt}
\end{equation}

\begin{equation}
n_{\rm s} -1 \doteq \frac{d\log\mathcal P_{\rm s}}{d\log k}=-2\epsilon_1-\epsilon_2.
\label{ns}
\end{equation}
In this subsection we want to estimate the impact of the scale dependence choosing a specific slow roll model of inflation, namely the Starobinsky model \cite{Starobinsky:1980te,Renzi:2019ewp} that predicts the following well known relations:
\begin{equation}
n_{\rm s}-1\simeq -\frac{2}{N} , \quad r\simeq\frac{12}{N^2}
\end{equation}
where $N$ is the e-fold number of inflation that we can fix since we measure $n_{\rm s}\simeq0.96$ with good precision \cite{Akrami:2018odb}.  The tensor tilt and its runnings read
\begin{equation} 
n_{\rm t}\simeq -\frac{3}{2}\left(\frac{1}{N}\right)^2 ; \quad  \alpha_n^t\simeq -\frac{3}{2} (n+1)! \left(\frac{1}{N}\right)^{n+2}
\end{equation}
The sum expansion that quantifies the scale dependence of the tensor tilt can be easily computed to be
\begin{equation}
\sum_{n=1}^{\infty} \frac{\alpha_n^{\rm t}(k_*)} {(n+1)!} \left[\log\left(\frac{k}{k_*}\right)\right]^n\simeq  n_{\rm t} \left(\frac{\frac{1}{N}\log(\frac{k}{k_*})}{1-\frac{1}{N}\log(\frac{k}{k_*})}\right)
\end{equation}

\begin{figure}[h]
	\centering
	\includegraphics[width=0.65\linewidth]{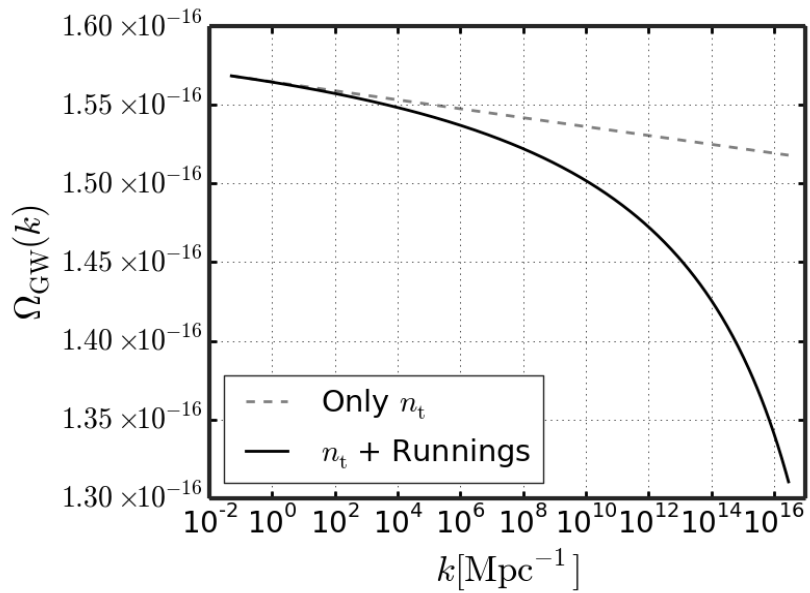}
	\caption{$\Omega_{\rm{GW}}(k)$ in the Starobinsky model both including (solid lines) and neglecting (dashed lines) the runnings. The scale dependence is negligible.}
	\label{fig:figure2}
\end{figure} 
In Fig. \ref{fig:figure2} we plot $\Omega_{\rm{GW}}(k)$ from the CMB scales all the way up to the GW scales both including (black solid line) and neglecting (gray dashed line) the runnings. As one can see the runnings lead to negligible corrections also on small scales. This is not surprising since by definitions the slow roll paradigm predicts an almost scale independent slightly red tilt. We actually study this model to provide an example of negligible scale dependence and to show how the situation can be drastically different in more elaborated scenarios as those discussed in the next subsection.

\subsection{Particle Production} \label{sec.ParticleProduction}
In this subsection we want to provide a counterexample studying a different physical model of inflation that employs a pseudo scalar axion naturally coupled to gauge fields. In this model a mechanism of particle production takes place during the rolling inflation and this can be translated into a blue spectrum of gravitational waves. We will show that the tensor tilt can acquire a non trivial scale dependence as well.
We start giving a brief description of the model, more details can be found in \cite{Anber:2009ua,Pajer:2013fsa,Mukohyama:2014gba,Namba:2015gja,Ferreira:2015omg,Peloso:2016gqs}. We consider a simple theory of a Pseudo Nambo Goldstone Boson inflation. In this model the inflaton field $\phi$ and the axion $\psi$ are minimally coupled to gravity and the axion is also coupled with a $U(1)$ gauge field in a way consistent with symmetries\footnote{Note that the axion is not the inflaton itself but another distinct field.}. The action of the theory is 
\begin{equation}
S=\int d^{4} x \sqrt{-g}\left[\frac{M_{p}^{2}}{2} R-\frac{1}{2}(\partial \phi)^{2}-V(\phi)-\frac{1}{2}(\partial \psi)^{2}-U(\psi)-\frac{1}{4} F^{\mu\nu}F_{\mu\nu}-\frac{\psi}{4 f} F_{\mu\nu} \tilde{F}^{\mu\nu}\right]
\end{equation}
$F_{\mu\nu}$ and $\tilde{F}^{\mu\nu}\doteq\frac{1}{2} \epsilon^{\mu \nu \alpha \beta} F_{\alpha \beta}$ are the field-strength tensor of the gauge field and its dual, respectively; $f$ is the axion decay constant while $V(\phi)$ and $U(\psi)$ are the inflation and axion potential. We also assume a flat FRW metric and that both the inflaton and the axion take a homogeneous vacuum expectation value (vev) while the gauge field carries no vev. Under this assumption the equations of motion for the inflaton and the axion are
\begin{equation}
\ddot{\bar{\phi}}+3 H \dot{\bar{\phi}}+V^{\prime}(\bar{\phi})=0
\end{equation}
\begin{equation}
\ddot{\bar{\psi}}+3 H \dot{\bar{\psi}}+U^{\prime}(\bar{\psi})=0
\end{equation}
where the prime denotes the derivatives with respect to the argument and the over-dots denote the derivatives with respect to time. We also assume that  the contribution of the axion on the background evolution is negligible compared to that of the inflaton i.e $\left| U \right|\ll V$ and $\dot{\bar{\psi}}^{2} \ll \dot{\bar{\phi}}^{2}$. We introduce the parameter 
\begin{equation}
\xi \equiv \frac{\dot{\bar{\psi}}}{2 H f}
\end{equation}
that will play a crucial role in our future discussion. We assume $\xi$ to be nearly but not exactly scale independent:
\begin{equation}
\xi_1\doteq\frac{d\log\xi}{d\log k}=\frac{\dot \xi}{\xi\,H}\ll 1.
\label{xi1}
\end{equation}
We instead  assume $\xi_1$ to be constant  i.e. $d\log\xi_1/d\log k\approx0$. In our future discussion we restrict our attention to the case $\xi>1$ that allows a blue tensor tilt. We are not going to discuss in details  the  peculiarities of this model  such as the gauge quanta production \cite{Anber:2009ua} that are reviewed also in \cite{Mukohyama:2014gba,Peloso:2016gqs} and the references within, but for our aim it is sufficient to observe that, in order to avoid a significant back-reaction of the produced gauge quanta to the background dynamics, we have to require that 
\begin{equation}
\frac{\mathrm{e}^{\pi \xi}}{\xi^{5 / 2}} \ll \frac{13.5}{\sqrt{\epsilon_1 \mathcal{P}_0}} \frac{f}{M_{p}}
\label{decay const}
\end{equation}
where $\mathcal P_0= \left(\frac{1}{8\pi^2 M_{p}^2}\right)\left(\frac{H^2}{\epsilon_1}\right)$ is the primordial scalar spectrum without source (i.e. as predicted by the slow roll inflation). The scalar and tensor spectra for this model are \cite{Barnaby:2012xt, Mukohyama:2014gba,Sorbo:2011rz}:
\begin{equation}
\mathcal P_{\rm s} \simeq \mathcal{P}_0\left(1+c_{\rm s}\,\epsilon_1^{2} \,\mathcal{P}_0 \,\frac{\mathrm{e}^{4 \pi \xi}}{\xi^{6}}\right)\bigg \rvert _{k=k_*}
\label{Ps}
\end{equation}
\begin{equation}
r\simeq\frac{16\, \epsilon_1 \,\left(1+c_{\rm t} \,\epsilon_1 \, \mathcal{P}_0 \frac{\mathrm{e}^{4 \pi \xi}}{\xi^{6}}\right)}{\left(1+c_{\rm s}\,\epsilon_1^{2} \,\mathcal{P}_0 \,\frac{\mathrm{e}^{4 \pi \xi}}{\xi^{6}}\right)} \bigg \rvert _{k=k_*}
\label{Pt}
\end{equation}
where $c_{\rm s}= 2.5 \cdot 10^{-6} $ and $c_{\rm t}=3.4 \cdot 10^{-5}$ are constants.
We compute the spectral tilts from the relation \eqref{Ps} and \eqref{Pt} taking the logarithm derivatives:
\begin{align}
n_{\rm s}-1\doteq \frac{d\log\mathcal P_{\rm s}}{d \log k}\bigg \rvert _{k=k_*}&=\frac{d\log\mathcal P_0}{d\log k} + \frac{c_{\rm s} } {1+c_{\rm s}\,\epsilon_1^2 \,\frac{\mathrm{e}^{4 \pi \xi}}{\xi^{6}}}\,\frac{d}{d\log k}\left(\epsilon_1^2 \mathcal P_0 \frac{\mathrm{e}^{4 \pi \xi}}{\xi^{6}} \right)\\
&=-2(1+f_{\rm s} )\epsilon_1-(1-f_{\rm s})\epsilon_2+f_{\rm s} (4\pi\xi-6)\xi_1\\
&\simeq -2\epsilon_1-\epsilon_2
\label{ns_blue} 
\end{align}
and 
\begin{align}
n_{\rm t}\doteq \frac{d\log\mathcal P_{\rm t}}{d \log k}\bigg \rvert _{k=k_*}&=\frac{d\log\epsilon_1}{d\log k} + \frac{d\log\mathcal P_0}{d\log k} + \frac{c_{\rm t} } {1+c_{\rm t}\,\epsilon_1 \,\frac{\mathrm{e}^{4 \pi \xi}}{\xi^{6}}}\,\frac{d}{d\log k}\left(\epsilon_1 \mathcal P_0 \frac{\mathrm{e}^{4 \pi \xi}}{\xi^{6}} \right)\\
&=-2(1+f_{\rm t})\epsilon_1+f_{\rm t}(4\pi\xi-6)\xi_1
\label{nt_blue}
\end{align}
where  the functions
\begin{equation}
f_{\rm s}\doteq \frac{c_{\rm s}\,\mathcal P_0\,\epsilon_1^2\,  \frac{\mathrm{e}^{4 \pi \xi}}{\xi^{6}} } {1+c_{\rm s}\,\mathcal P_0\,\epsilon_1^2\,  \frac{\mathrm{e}^{4 \pi \xi}}{\xi^{6}} }\ll 1
\label{fs}
\end{equation}
and
\begin{equation}
f_{\rm t}\doteq\frac{c_{\rm t}\,\mathcal P_0\,\epsilon_1\,  \frac{\mathrm{e}^{4 \pi \xi}}{\xi^{6}} }{1+c_{\rm t}\,\mathcal P_0\,\epsilon_1\,  \frac{\mathrm{e}^{4 \pi \xi}}{\xi^{6}} }
\label{ft}
\end{equation}
weigh the corrections to the slow roll predictions for the scalar and tensor parameters respectively. 
In what follows we fix $\mathcal P_{s}$ and $n_s$ to the observed values $\mathcal{P}_{\rm s}\simeq 2.1\times 10^{-9}$ and $n_{\rm s}\simeq 0.96$ \cite{Akrami:2018odb}. We also fix the tensor to scalar ratio $r$ to reference value $r\simeq 10^{-2}$ and $\xi_1\simeq 5 \times 10^{-3}\ll1$. Note that our results are marginally sensitive to the value of $r$ and $\xi_1$ and that we are not interested into a parameter analysis for this specific model: our task is simply to show that also in physical models of inflation scale dependence can be non-negligible.  

\begin{figure}[h]
	\centering
	\includegraphics[width=0.9\linewidth]{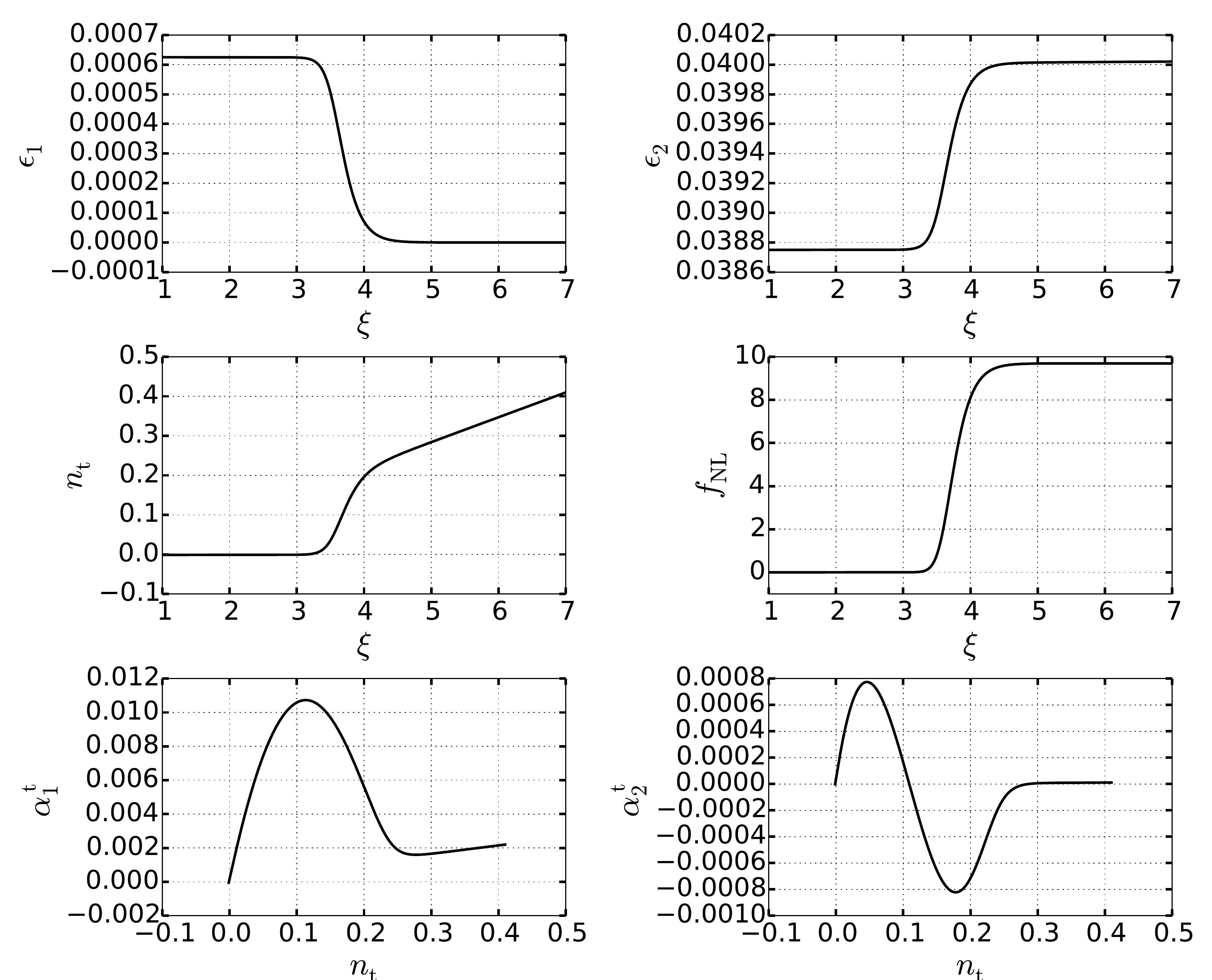}
	\caption{The parameters of the model as functions of $\xi$.}
	\label{fig:figure3}
\end{figure}

We use the Eqs. \eqref{Ps} and \eqref{Pt} in order to explicit $\epsilon_1$ and $\mathcal P_0$ as functions of $\xi$. This means that when $\xi$ changes, $\epsilon_1(\xi)$ and $\mathcal P_0(\xi)$ change in such a way that $\mathcal P_s$ and $r$  remain constant. Moreover because of Eq. \eqref{nt_blue} also $n_{\rm t}$ is only a function of $\xi$. Being $n_{\rm s}$ fixed by observations, we can also use the relation \eqref{ns_blue} in order to find $\epsilon_2$ as a function of $\xi$ so that when $\xi$ changes, $\epsilon_2(\xi)$ changes leaving $n_{\rm s}$ fixed to its observed value. So in this model all the inflationary parameters\footnote{The inflationary parameters are to be considered evaluated to the pivot scale $k=k_{*}=0.05\rm{ Mpc}^{-1}$ which means that also the parameter $\xi$ in the equations above is computed on the CMB scales $\xi=\xi(k=k_*)$.} become known functions of $\xi$. We plot them in Fig. \ref{fig:figure3} letting $\xi$ vary in the range $\xi\in[1\,,\,7]$. 

First of all we want to stress that we have carefully checked that the scalar spectrum \eqref{Ps} remains essentially equal to $\mathcal P_0$ (that is what predicted by the single field slow roll inflation). As a matter of fact,  if we decompose the scalar spectrum $\mathcal P_{\rm s}=\mathcal P_0 + \mathcal P_{\rm{s,\,sourced}}$ we find that  the sourced term induces corrections that are extremely small compared to the vacuum contribution $\mathcal P_{\rm{s,\,sourced}}\sim 10^{-4}\,\mathcal P_0$ for all the values of $\xi$. In other words, the corrections to the scalar spectrum are completely negligible ($f_{\rm s}\approx 0$), and the scalar parameters are essentially equal to that obtained in the simplest slow roll models. This can be understood by noting that the scalar corrections are suppressed by a factor $\epsilon_1^2\,\mathcal P_0$ and that $\epsilon_1$ exponentially decreases with $\xi$ in order to keep $r$ fixed, see also Fig.  \ref{fig:figure3}.  The fact that the scalar spectrum is essentially indistinguishable from the single field slow roll models is crucial since in this way all the tight constraints on the scalar perturbations (e.g. their high level of gaussianity) are respected as well \cite{Barnaby:2012xt,Shiraishi:2019yux}.  
On the other hand the corrections to the tensor spectrum can be dominant for an appreciable range of the parameter space, allowing also a blue tensor tilt, see Fig. \ref{fig:figure3}. The sourced tensor modes could also leave a sizable non-gaussianity of nearly equilateral shape on the CMB anisotropies and polarization. The amount of non-gaussianity is controlled by the parameter $f_{\rm{NL}}$ estimated as \cite{Mukohyama:2014gba,Cook:2013xea}: 
\begin{equation}
f_{\rm{NL}}\simeq 1.1\times 10^{-14}\left(\epsilon_1\,\frac{e^{2\pi\xi} }{\xi^3}\right)^3
\end{equation}
and its shape given in Fig. \ref{fig:figure3}, as well.
We estimate the scale dependence of the tensor tilt performing a second order computation and deriving the expression for the tensor running $\alpha_1^{\rm t}\doteq dn_{\rm t}/d\log k$ and the running of the running $\alpha_2^{\rm t}\doteq d\alpha_1^{\rm t} / d\log k$:
\begin{equation}
\alpha_1^{\rm t}\doteq\frac{dn_{\rm t}}{d\log k}\bigg \rvert _{k=k_*}=-2(1+f_{\rm t})\epsilon_1\epsilon_2 -2 f^{\prime}_{\rm t}\epsilon_1 +f^{\prime}_{\rm t}(4\pi\xi-6)\xi_1 +4\pi f_{\rm t} \xi\xi_1^2 
\label{alpha_1}
\end{equation}
\begin{equation}
\alpha_2^{\rm t}\doteq\frac{d\alpha_1^{\rm t}}{d\log k}\bigg \rvert _{k=k_*}=-2(1+f_{\rm t})\left(\epsilon_1\epsilon_2^2+\epsilon_1\epsilon_2\epsilon_3\right)-4 f^{\prime}_{\rm t}\epsilon_1\epsilon_2-2 f^{\prime \prime}_{\rm t}\epsilon_1 +f^{\prime \prime}_{\rm t} (4\pi\xi-6)\xi_1 +8\pi f^{\prime}_{\rm t}\xi\xi_1^2 +4\pi f_{\rm t} \xi \xi_1^3
\label{alpha_2}
\end{equation}
where we have defined:
\begin{equation}
f^{\prime}_{\rm t}\doteq\frac{d f_{\rm t}}{d\log k}\bigg \rvert _{k=k_*}=\left[ \frac{-2\epsilon_1 + (4\pi\xi - 6)\xi_1}{1+c_{\rm t}\mathcal P_{0} \epsilon_1 \frac{\mathrm{e}^{4 \pi \xi}}{\xi^{6}}  } \right] f_{\rm t}
\end{equation}
and
\begin{equation}
\begin{aligned}
f^{\prime \prime}_{\rm t}\doteq \frac{df^{\prime}_{\rm t}}{d\log k}\bigg \rvert _{k=k_*} =&\left[ \frac{-2\epsilon_1 + (4\pi\xi - 6)\xi_1} {1+c_{\rm t}\mathcal P_{0} \epsilon_1 \frac{\mathrm{e}^{4 \pi \xi}}{\xi^{6}}  } \right]^2 f_{\rm t}+\\
&+\left[ \frac{(1+c_{\rm t} \mathcal P_0 \epsilon_1 \frac{\mathrm{e}^{4 \pi \xi}}{\xi^{6}} ) (-2\epsilon_1\epsilon_2+4\pi\xi\xi_1^2) - c_{\rm t} \mathcal P_0 \epsilon_1 \frac{\mathrm{e}^{4 \pi \xi}}{\xi^{6}}\left[-2\epsilon_1+(4\pi\xi-6)\xi_1\right]^2} {\left(1+c_{\rm t}\mathcal P_{0} \epsilon_1 \frac{\mathrm{e}^{4 \pi \xi}}{\xi^{6}}\right)^2  } \right]f_{\rm t}
\end{aligned}
\end{equation}

In this model the tensor tilt can acquire a non trivial scale dependence. In fact, depending on the parameters, $d\log n_{\rm t}/d\log k \simeq 0.1$, see Fig. \ref{fig:figure3}. As we discussed in Sec. \ref{sec.ScaleDependence}, this can lead to non negligible corrections on small scales.

\begin{figure}[ht!]
	\centering
	\includegraphics[width=0.65\linewidth]{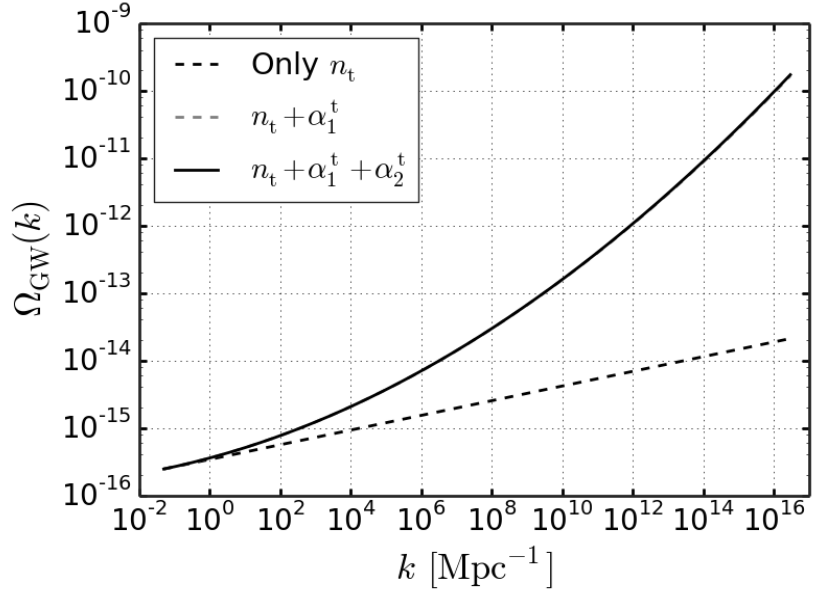}
	\caption{$\Omega_{\rm{GW}}(k)$ in the particle production model both including and neglecting the first two runnings. The scale dependence is not negligible.}
	\label{fig:figure4}
\end{figure}

As explained before all these quantities are known functions\footnote{ Note that we parametrized the slow parameter $\epsilon_3$ appearing in \eqref{alpha_2} as $\epsilon_3=\gamma\,\epsilon_2$. Letting $\gamma$ vary in a range $\gamma\in\left[-1,1\right]$ no significant changes in $\alpha_2^{\rm t}$ are observed. We therefore fixed $\epsilon_3\simeq 0$.} of $\xi$ or equivalently $n_{\rm t}$. However, since for large values $\xi\gtrsim 5$ the backreaction becomes typically non negligible as well as the primordial non gaussianity, we decide to restrict our attention to a safer region of the parameter space. We therefore fix  $n_{\rm t}\simeq0.1$ (or equivalently $\xi\simeq 3.5$) in such a way that both backreaction and non gaussianity are still under control, see Fig. \ref{fig:figure3}.  In this way the running  $\alpha_1^{\rm t}\simeq 0.01$ and the running of running $\alpha_2^{\rm t}\simeq 3\times 10^{-6}$ are fixed as well. We let evolve $\Omega_{\rm GW}(k)$ from the CMB scales all the way up to the ultra high $k$ probed by the ground based interferometers both including and neglecting $ \alpha_1^{\rm t}$ and $\alpha_2^{\rm t}$, see Fig. \ref{fig:figure4}.  The importance of scale dependence in this model is evident as $\Omega_{\mathrm{GW}}$ differs by many orders of magnitude when the non-linear corrections are considered, possibly becoming visible to future gravitational wave experiments such as LISA \cite{Audley:2017drz} and Einstein Telescope \cite{Punturo:2010zz}. 

While we have shown that the impact of the second order running $\alpha_2^{\rm t}$ is completely negligible, see also Fig.\ref{fig:figure4},  one may ask if the higher order terms $\alpha^{\rm t}_{n>2}$ can instead give an appreciable contribution possibly changing the shape of $\Omega_{\rm GW} (k)$. For our aim it is sufficient to note that being the tensor tilt only a function of $\xi$, the derivative with respect to the scale can be written as $d/d\log k=\left(d\xi/d\log k\right) \,d/d\xi=\left(\xi\xi_1\right) d/d\xi$ and that the overall factor $\xi \xi_1\sim 10^{-2}$ will further suppress the higher order derivatives. We therefore expect such terms to be smaller and smaller at least in this range of the parameters space. We leave the detailed analysis of the sum expansion convergence suitable for future works.

\section{Conclusion} \label{sec.Conclusion}
Along with the  B-modes polarization, a satiable amount of primordial gravitational waves may have imprinted also the stochastic background of gravitational waves.
Being the small scale experiments on gravitational waves such as LIGO and VIRGO (and in future LISA or ET) sensitive to the stochastic background, they can be used together with the CMB data to put constraints on the inflationary parameters. In particular from the LIGO/VIRGO bound on $\Omega_{\rm GW}$, Eq.\eqref{LVlimit},  under the assumption of a scale independent tilt, an upper bound $n_{\rm t}<0.53$ at 95\% C.L. is derived combining the CMB and GW data \cite{Akrami:2018odb}. 
In this paper we focused on the robustness of assuming the tensor tilt as scale independent over a range of about eighteen order of magnitude $k\in\left[0.05 \,,\,\sim10^{16}\right]\,	\rm{Mpc}^{-1}$.  We have shown that when the assumption of scale independence is relaxed, the constraints may become very sensitive to the higher order corrections in the power law expansion and that even a tiny scale dependence can significantly change the above mentioned constraints. Due to the huge distance between the scales probed by the CMB and the GW interferometers, also a relatively small departure ($\sim 4\%$) from scale independence can significantly affect the shape of primordial spectrum possibly leading to non-negligible corrections. We concluded that scale dependence cannot always be ignored when constraints on the inflationary parameters are derived  from small scales data. We have also provided two examples in completely different physical models of inflation. We have first considered  the Starobisnky model showing that the scale dependence is negligible as one expects in the simplest slow roll paradigm. The Starobinsky model (and in general the standard slow roll models) provides an example where the non-linear corrections are typically negligible even on the ultra high $k$ probed by the ground based interferometers. However this is not always true. As a counterexample, we have analyzed a physical model of blue inflation which employs a pseudo scalar axion naturally coupled to gauge fields showing that in this model the tensor tilt can acquire a non trivial scale dependence leading to non negligible corrections on small scales.

\section*{Acknowledgments}
WG and AM thank Antonio Riotto for the useful collaboration and for his precious suggestions that contributed to the realization of this article. WG thanks Marco Peloso and Fabrizio Renzi for their comments and valid clarifications. AM and WG are supported by "Theoretical Astroparticle Physics" (TAsP), iniziativa specifica INFN.

\bibliography{main.bib}
\end{document}